# Optimized conditions for direct imaging of bonding charge density in electron microscopy

J. Ciston[1,2*], J.S. Kim[3], S.J. Haigh[3], A.I. Kirkland[3], L.D. Marks[2]

[1] Center for Functional Nanomaterials, Brookhaven National Laboratory, USA

[2] Department of Materials Science and Engineering, Northwestern University, USA

[3] Department of Materials, University of Oxford, UK

**Abstract**

We report on the observability of valence bonding effects in aberration-corrected high resolution electron microscopy (HREM) images along the [010] projection of the mineral Forsterite ($Mg_2SiO_4$). We have also performed exit wave restorations using simulated noisy images and have determined that both the intensities of individual images and the modulus of the restored complex exit wave are most sensitive to bonding effects at a level of 25% for moderately thick samples of 20-25 nm. This relatively large thickness is due to dynamical amplification of bonding contrast arising from partial de-channeling of 1s states. Simulations also suggest that bonding contrast is similarly high for an un-corrected conventional electron microscope, implying an experimental limitation of signal to noise ratio rather than spatial resolution.

# 1. Introduction

In recent years, much effort has been devoted toward the goal of stable refinement of the subtle perturbations to valence charge density of both bulk [1, 2] and surface [3] structures against experimental electron diffraction data. The ability to reliably measure subtle bonding effects experimentally is especially critical for materials with strongly correlated electrons, such as high-Tc superconductors, multiferroics, and oxide catalysts, which can present computational difficulties with even the most sophisticated density functional theory (DFT) methods. Although these experimental efforts have been relatively successful, the non-convexity of the phase problem in diffraction makes the process of uniquely ascribing bonding features responsible for perturbations to measured diffraction intensities precarious. In addition, diffraction may only be used to obtain details of the bonding in perfect crystals, and provides no information about defect sites which often govern the performance of a material. Therefore, the measurement of local crystallographic perturbations to charge redistribution would be more useful if it were possible using direct imaging in real-space.

A prior study by Deng and Marks [4] surveying a wide variety of light-element oxides suggested that the mineral Forsterite ($Mg_2SiO_4$) exhibits relatively high sensitivity to bonding effects in simulated HREM images. Sensitivity to bonding along the [010] direction (equivalent to [100] in Deng's notation) was concluded to be 14% of the total image contrast under chromatic aberration (Cc)-limited imaging conditions ($C_3 = 0.005$mm) and 23% under fifth order spherical aberration ($C_5$)-limited conditions ($C_3 = -0.005$mm).)[5] The observability of bonding effects in HREM images was calculated to be even greater, as large as 50% of the total contrast, for charge defects at the (111) surface of MgO [6], although it should be noted that this large effect was likely due to the use of an incorrect structure that was not valence-compensated through surface hydroxylation [7].  There were two areas in which these prior analyses was lacking.  Although a wide variety of light element oxide materials were studied, quantification was performed only at the imaging conditions selected for maximum qualitative interpretability in the simulated images.  Consequently, all estimates of the observability of bonding effects in [4] were only performed for samples at a thickness of 5nm and defocus of ±35 Å, which are not necessarily the parameters which maximize the contribution of bonding effects to the image.  The second deficiency in previous studies was the lack of incorporation of the effects of detector noise on the practical observability of these bonding perturbations in experimental images.

In this paper, results are presented on the observability of valence bonding effects in aberration-corrected high resolution electron microscopy (HREM) images along the [010] projection of the mineral Forsterite ($Mg_2SiO_4$).  The first aim of the work presented herein was to more fully explore the thickness and defocus parameter space for the detection of the effects of bonding in simulated images.  Our simulations also incorporated an estimation of experimental detector noise.  In addition, we have performed exit wave restorations using a series of simulated noisy images and have determined that the primary advantage of the exit wave technique is not due to the increase in image resolution, but a reduction of noise to acceptable levels for the observation of subtle bonding effects.  The motivation for this computational analysis was the determination of experimental conditions under which bonding effects are most likely to be observed.

## 2. Computational methods

*2.1. Charge Density Multislice Method*

Simulation of both images and dynamical diffraction patterns were accomplished using the well-established multislice method [8, 9] as implemented in the NUMIS 2.0 code at Northwestern University [10]. Traditional multislice simulations calculate the phase grating of each slice by summing isolated-atom electron scattering factors (IAM) for atoms at crystallographic positions. This method is convenient because it a) allows the discrete partitioning of whole atoms into slices b) the atomic scattering factors can be computed once and subsequently used for all possible structures and c) the integration of the potential is reduced to a finite coherent sum of the individual scattering factors convolved with the structural lattice. However, IAM scattering factors are not sufficient for the description of the subtle perturbations to the potential due to bonding effects. Two important problems arise when one considers the possibility of performing multislice calculations that incorporate bonding effects, namely the *a-priori* determination of the valence charge density, and the correct way to partition a continuous charge density into discrete potential phase gratings.

The method first implemented by Deng *et al* [4, 6] solves both of these problems in a relatively user-friendly fashion. The first step towards full charge density multislice was the calculation of the self-consistent charge density utilizing Density Functional Theory (DFT) methods as described below. It is important that the results from DFT calculations are well converged because small perturbations to the charge density can substantially affect the electrostatic potential due to modifications to the screening of the positive atomic cores. If the DFT calculations are performed using an all-electron code such as Wien2k [11], X-ray structure factors may be easily computed as a Fourier transform of the total charge density. This method is computationally less intensive than using the real-space potential directly due to the fine voxel sampling required to describe the core regions and discontinuities at the muffin tin edges. These X-ray structure factors ($F^x$) can be subsequently converted to electron structure factors ($F^e$) using the Mott-Bethe formula

$$F^e(h',k',l') = \frac{2m_0 e^2}{h^2} \frac{(Z - F^x(h',k',l'))}{s^2} \exp(-B \cdot s^2),$$
$$Z = \sum_i Z_i \cdot \exp(i2\pi(x_i \cdot h + y_i \cdot k + z_i \cdot l))$$

(1)

where B is the Debye-Waller factor, s is the magnitude of the scattering vector, ($x_i$, $y_i$, $z_i$) are atomic coordinates of the $i^{th}$ atom, and $Z_i$ is the atomic number of the $i^{th}$ atom. In the calculations reported here, all Debye-Waller factors were set to a fixed isotropic value of 0.25 Å$^2$ since a beam-by-beam application of the Debye-Waller factor in the Mott-Bethe formula is not easily modified to accommodate individual Debye-Waller factors for each atom. However, the errors incurred by the use of a single Debye-Waller factor are minimized for the low-angle reflections which are most sensitive to charge density variations. The reciprocal space phase grating, V'(h,k), was calculated by projecting $F^e$ along the [010] direction by the simple relationship, V'(h,k) = $F^e$(h',0,l'). To allow direct comparison of the charge density results to those from a traditional neutral-atom multislice, the amplitude of the phase grating was divided by an integer number of slices per unit cell to reproduce the slice thickness in comparable traditional multislice calculations.

It is important to note that this method creates a two dimensional potential projected through the full unit cell rather than allowing for differing numbers of atoms in each layer. While this method of handling the potential ignores higher-order Laue zone (HOLZ) reflections, HOLZ diffraction occurs predominantly at large angles and we can expect the effects to be small for the low-angle reflections most sensitive to bonding perturbations. Moreover, it has been shown previously that the differences between a traditional IAM multislice and the use of the Deng projection method for a superposition of IAM densities are small [4].

*2.2. HREM Image Simulations*

Multislice image simulations were carried out along the [010] direction of the mineral Forsterite (Mg$_2$SiO$_4$, a=0.4752 b=1.0193 c=0.5997 nm, Pbnm [12] – see structure file in CIF format in supplementary materials). Forsterite was chosen as the initial test case due to its good performance in prior work screening multiple oxide materials [4], and also for its lack of d-orbitals or other electronic effects that may complicate the DFT calculations. Therefore, it is assumed that the DFT charge density is a robust representation of the true charge density. Projected potentials were derived from two sets of X-ray structure factors. The first set of X-ray structure factors was calculated from a linear superposition of Dirac-Slater orbitals representing the isolated atom (IAM) charge densities calculated using a relativistic Hartree-Fock approach [13] as implemented in the Wien2k program [11]. The second set was derived from the full charge density of the unit cell (CD) as calculated by the all-electron full-potential Wien2k code. The DFT calculation was performed spin-unpolarized utilizing the PBE-GGA functional [14] with muffin-tin radii for all atoms fixed at 1.5 a.u., RKMAX= 7.0, and a 10x4x8 k-point mesh. Monolayer BN was computed in an orthorhombic supercell (a=1.0249 b=0.2514 c=0.4354 nm) with a 4 atom basis and Pmm2 symmetry, RKMAX = 7.5, and a 2x8x5 k-point mesh. X-ray structure factors were calculated as a Fourier transform of the fully converged charge density.

Both sets of X-ray structure factors (IAM and CD) were converted to electron scattering factors via the Mott-Bethe formula with temperature factors of B=0.25 Å$^2$ used for all atoms which were then projected along the [010] direction of the unit cell. The projected potential was sliced into 6 layers each with an identical potential of $V_{total}$/6 and a thickness of 0.17nm.

Images were calculated utilizing the parameters of the spherical aberration corrected JEOL JEM-2200MCO microscope [15] under two limiting cases: $C_5$-limited with third order spherical aberration, $C_3$= -0.005 mm and $C_c$-limited with third order spherical aberration $C_3$=+0.005 mm [5]. Other microscope parameters used were an accelerating voltage of 200kV, beam convergence of 0.1 mrad, $C_c$=1.2 mm, $\Delta V_{acc}/V_{acc}$= 0.3 ppm, $\Delta I_{obj}/I_{obj}$,= 0.5 ppm, $\Delta E$=0.5 eV. We made no attempt to incorporate contributions from inelastic scattering in this analysis due to the light elements and thin samples considered in these calculations, and the wide availability of energy filters on modern microscopes. Poisson noise was added to the simulated images at a level of 10% of the total contrast to simulate experimental conditions. While this is a higher level of noise than the 3-5% value which would be expected from pure "shot noise" for typical illumination (500-2000 counts/pixel), additional noise was added to approximate the effects of image degradation due to surface contamination / damage, crystal defects, etc. Figure 1 shows a collage of simulated images along the [010] projection of Forsterite at a variety of thickness and defocus values simulated using the fully-bonded CD potential with a $C_3$ value of -0.005mm. Images calculated for identical microscope parameters using the IAM potential appear visually identical and are therefore not shown. Figure 2 shows the difference between images simulated with IAM and CD potentials for the same microscope conditions which will be discussed in section 3.

Exit wave restorations were performed on each focal series of simulated images using a linear Wiener filter [16-19]. Restoration of the complex exit wave has two primary benefits: firstly, the phase of the complex exit wave is often, though not always, more directly interpretable than an individual image since the reversals in the contrast transfer function have been removed; secondly, the restored exit wave has a better signal to noise ratio than any individual image. In this work, exit waves were computed for both clean and noisy simulated images for $C_3 \pm 0.005$mm. Fourteen images were used in each focal series restoration with defocus values from -35nm to 30nm in 5nm increments. Difference maps of both the modulus and phase components of the restored exit waves were also calculated (as IAM subtracted from CD, ((CD-IAM)) ) to gain insight into the specific features most sensitive to bonding. Figure 3 shows a collage of the exit wave modului and phases restored from simulated focal series of Forsterite samples with varying thickness and a $C_3$ value of -0.005 mm for both IAM and CD potentials.

*2.3. Quantification of contrast attributable to bonding*

In order to compare the findings of this study to previously published results, a quantitative measure of the image contrast due to bonding based upon standard uncertainty of global pixel intensity in a gray scale image ($R_\sigma$) was calculated for each IAM/CD pair by comparing difference maps to IAM simulations. We have utilized the figure of merit given in [4] calculated as:

$$R_\sigma = \frac{\sigma_{(CD-IAM)}}{\sigma_{IAM}} \qquad (2)$$

Where, σ is the standard deviation of the total contrast of a CD, IAM, or difference (CD-IAM) image. The $R_\sigma$ metric has the advantage of being simply interpreted as the fractional contrast due to bonding. However, this figure of merit tends to yield unrealistically large values in cases where the total contrast is small, such as for very thin samples and Gaussian defocus images. While this effect is not, in principle, problematic for a computational study, in real experiments the low-contrast images would be more challenging to interpret due to signal/noise limitations. To compensate for this, we have used an intensity-scaled pixel-by-pixel $R_1$ metric which is more appropriate, being more sensitive to sharply peaked differences between the CD and IAM images most relevant to experimental observation. The pixel-by-pixel evaluation is well suited for this study, however when comparing to experimental data, slight variations in sampling interval or global blurring may lead to unusually large difference values. The $R_1$ figure of merit is defined as:

$$R_1(CD, IAM) = \frac{\sum_{pixels} |s*CD_i - IAM_i|}{\sum_{pixels} |IAM_i|} \qquad (3)$$

where s is a scaling term utilized for intensity conservation given by:

$$s(CD, IAM) = \frac{\sum_{pixels} |CD_i * IAM_i|}{\sum_{pixels} |IAM_i|^2} \qquad (4)$$

To determine the robustness of the observable bonding contribution in noisy images, we used a modified $R_1$-type "observability index" ($R_{1,obs}$) which effectively weights the images based upon their signal to noise ratio:

$$R_{1,obs} = R_1(CD, IAM) * [1 - R_1(clean, noisy)] \qquad (5)$$

where $R_1(clean,noisy)$ takes the same form as equation (3) comparing clean and noisy CD images rather than CD and IAM images. For a fully uncorrelated noisy image, $R_1(clean,noisy)$ will approach unity, leading to a $R_{1,obs}$ value of zero.

### 3. Simulation Results

*3.1. Qualitative image interpretation*

We begin our analysis with a qualitative correlation of specific features in the simulated images with details of the Forsterite structure. It is clear from Figure 1 that many of the simulated images exhibit one pair of bright features per unit cell. A more detailed view of one image in the series (15 nm thick, +10 nm defocus) is shown in Figure 4a with the bulk unit cell overlaid. From the outset it is immediately clear that image interpretation is difficult even when using aberration-corrected optics. For example, in Figure 4a, the silicon atoms and their bridging oxygen bonds appear as bright features while all of the magnesium atoms appear as dark features. If the structure was not known *a-priori*, it would be very difficult to assign the atomic positions simply by visual inspection of this image.

Three classes of features may be distinguished from the detailed view of an individual CD-IAM difference map as shown in Figure 4b. Firstly, the region near the silicon atoms in projection exhibits an excess of intensity when bonding effects are taken into account. Secondly, the region near the bridging oxygen atoms in projection exhibits a depletion of intensity when bonding effects are taken into account. Thirdly, the isolated magnesium atoms as well as the non-silicon-bridging oxygen atoms are largely unaffected by the incorporation of bonding effects and appear grey. The disparity between the bonding contrasts due to magnesium atoms projected in the same columns as silicon atoms against the isolated magnesium columns implies that the bright bonding features were related specifically to silicon. This analysis also indicates that the bonding contrast in HREM images is more sensitive to covalent effects than ionic effects as Si-O bridging bonds are expected to be more covalent than the Mg-O bridging bonds. It is important to note that the Si-O bridging feature, with the largest contribution to the bonding contrast, has a periodicity of 2.4 Å (spatial frequency of 0.42 Å$^{-1}$). This is consistent with the findings of Deng *et al* [4] in which the maximum deviation of the electron structure factors due to bonding effects was calculated to be in the spatial frequency range 0.2-0.4 Å$^{-1}$.

Figure 5 shows detailed view of the restored modulus of the exit wave from a simulated focal series of a Forsterite crystal identical in thickness to that shown in Figure 4. Whilst the details are somewhat different, our primary interpretation is almost identical with silicon atoms shown as bright features in both the full image (5a) and the CD-IAM difference map (5b), and Si-bridging oxygen atoms exhibiting charge depletion in the difference map. However, when the sample thickness used for simulation was increased from 15nm to 25 nm, the CD-IAM difference map of the exit wave modulus exhibited dark features near all of the oxygen atoms indicating that for thicker samples, ionic bonding effects become equally important to covalent effects.

*3.2. Analysis of quantitative bonding contrast*

For this analysis to be useful in guiding future experimental studies, the first question which must be answered concerns the appropriate value of $C_3$ to be used for observation of valence effects. Figure 7 is a thickness/defocus plot of the $R_\sigma$ metric for a) $C_3 = -0.005$ mm, b) $C_3 = +0.005$ mm, and c) the difference between $+C_3$ and $-C_3$ conditions. The periodicity of the vertical bands of high charge contrast is related to the sweeping of nodes and anti-nodes of the contrast transfer function through the charge-sensitive 0.2-0.4 Å$^{-1}$ spatial frequencies as the defocus was varied. Within this the global parameter space the sensitivity to bonding effects was similar at positive and negative $C_3$ values with negative $C_3$ being on average slightly superior. Although a small negative $C_3$ serves to partially compensate for uncorrected $C_5$, the relative insensitivity of bonding contrast to the precise value of $C_3$ is a consequence of the long range character of valence electrons, which are largely unaffected by small improvements to the resolution of high spatial frequencies.

When the $R_1$ metric for bonding contributions was applied to the simulated images of Forsterite (Figure 8a-c), the vertical banding became less obvious and the maximum sensitivity to bonding was achieved for a 20nm thick sample at -15 nm defocus, yielding an $R_1$ value of 32.9%. Figure 8c again indicated that a negative $C_3$ value was slightly superior and thus the remainder of the bonding contrast analysis reported here will be limited to negative values of $C_3$. The simulations were repeated for an accelerating voltage of 80 kV (Figure 9a) with the microscope parameters scaled appropriately, resulting in a larger effect due to chromatic aberration. As expected, the maxima of bonding observability occur at smaller thickness values. However, the overall contribution of bonding to the image contrast also has a smaller average value compared to 200 kV.

The spatial frequencies which are most sensitive to bonding contrast in Forsterite [010] correspond to real space periodicities of ~2.4 Å, which should be within the accessible range of a conventional TEM without aberration correction. Therefore, images were also simulated for a 200kV microscope with 1 mm $C_3$, and all other parameters identical to the JEOL 2200-MCO microscope used herein. Defocus values were scaled to maintain the same fractions of Scherzer defocus used for the aberration-corrected case. The $R_1$ values for these C3 = 1 mm simulations are similar to the $C_3$ = 0.005 mm case, and are shown in Figure 9b. While the R1 values are of similar magnitude for $C_3$ = 0.005 and 1 mm, the defocus bands where contrast is maximized are smaller for the conventional microscope and may require concomitantly finer defocus sampling to ensure that the optimal imaging conditions are experimentally explored.

Thus far the quantitative bond contrast analysis has focused on simulated images without consideration of experimental noise. The noise contribution ($R_1$(*clean,noisy*)) is plotted in Figure 10a and the total $R_{1,obs}$ is shown in Figure 10b. As expected, simulated images of crystals with lower thicknesses were more affected by the introduction of noise due to their overall weaker contrast. Exit waves were restored from the simulated images to determine whether this process is advantageous for imaging valence contributions compared to the use of individual images. The $R_1$ metric was applied to the moduli and phases of the restored exit waves for comparison to the individual thickness/defocus images discussed previously and the results are plotted in Figure

11. For purposes of comparison, the best and worst achievable bonding sensitivity for single images at each thickness (from Figure 8b) are also shown with the results from the exit waves.

For images simulated without experimental noise (Figure 11a), the restored exit wave moduli were globally more sensitive to bonding effects than their complement phases for all reasonable sample thicknesses (i.e. > 5nm).  This was somewhat unexpected since for thin crystals, the modulation of the phase of the outgoing electron wave typically has a greater effect on the overall image than modulus.  In addition neither the modului (31.1% $R_1$ at 25 nm thickness) nor the phases (9.2% $R_1$ at 15 nm thickness) of the restored exit waves were, in general, more sensitive to bonding effects than the best individual images attainable at specific defocus values (32.9% $R_1$ at 20nm thickness and -15nm defocus).  However, if the experimental defocus were not sufficiently sampled in an experiment, the specific combinations of thickness and defocus required to produce maximum bonding contrast may be missed in individual images.  Therefore, while the restored exit waves are not necessarily more sensitive to bonding than specific individual images, their use should be more robust due to a decreased reliance on precise defocus / thickness combinations that may not be experimentally accessible.

When noise was included in the images of the focal series and the exit wave was restored from this noisy data, the conclusion is different.  In general, when an exit wave is restored from a noisy series of images, the noise is reduced by a factor of $\sqrt{N}$, where N is the number of images in the series due to improved Poisson counting statistics.  In Figure 11b we plot $R_{1,obs}$ of the exit wave for each thickness together with the best and worst values from single images at each thickness (single image values taken from Figure 10b).  The $R_{1,obs}$ value for the restored exit wave modulus peaks at a thickness of 25nm ($R_1$=25.9%) which is the only value where the bond observability in the modulus has a greater value than a specific optimum individual image

(R1=13.9%). Furthermore, that the sensitivity of the exit wave to bonding peaks at the same thickness where the strongly ionic oxygen atoms became apparent in the modulus image shown in Figure 6.

*3.3. On the dynamical amplification of bonding contrast*

In order to understand the necessity of moderately thick samples for the robust detection of valence charge density, it is useful to turn to electron channeling theory [20-23]. For samples of intermediate thickness (>5 nm), the two-dimensional channeling eigenstates are dominated by deep 1s-symmetry eigenstates centered on the atomic cores [24, 25]. For crystallographic orientations that project well (i.e. have a small degree of overlap in the atomic potentials), a simplification to channeling theory may be made in which the wavefunction, $\psi(\mathbf{r},z)$, may be represented by a finite sum of 1s-type states centered at the atomic positions rather than the traditional infinite sum over periodic eigenfunctions [26] as:

$$\psi(\mathbf{R},z) - 1 = \sum_i C_i \phi_i(\mathbf{R} - \mathbf{R}_i) \left[ \exp\left(-i\pi \frac{E_i}{E_0} \frac{z}{\lambda}\right) - 1 \right] \qquad (6)$$

where $\varphi_i$ is the 1s channeling eigenstate for the *i*th atom centered at a vector $\mathbf{R}_i$ from the origin; z is the sample thickness, $E_0$ is the incident beam energy, and $E_i$ is the energy of the atomic eigenstate. The rate of change of the phase of the electron wave with thickness may thus be used to measure the energy of the dominant eigenstate.

The function $\psi(\mathbf{r})-1$ is plotted as an Argand diagram in Figure 12 for Forsterite [010] IAM and CD multislice simulations at thicknesses of 3.3 to 30 nm. For sample thicknesses less than 17 nm, the CD and IAM wavefunctions overlap almost completely. This is in agreement with the low bonding sensitivity of exit waves at low thicknesses restored from a series of simulated images as shown in Figure 11. At a sample thickness of 24nm (near the highest sensitivity to bonding electrons), the phase of the primary arm of the full CD electron wavefunction lags the IAM wavefunction by approximately 5 degrees on the argand diagram. This corresponds to an atomic eigenstate that is 0.5eV lower for the CD case than for the IAM. Because the eigenstates for the fully bonded case are very similar in energy to the IAM model, a relatively thick sample is required to accumulate enough dynamical amplification of this subtle effect to be observable in an HREM image.

To further validate this dynamical amplification result, we have compared 20 nm of Forsterite [010] to the extreme case of a thin, light element sample represented by a monolayer of boron nitride [27, 28]. BN was chosen for its high fraction of total electrons contributing to bonding, which would be expected to maximize the difference between images simulated with the IAM model against the full CD formalism. However, for a single monolayer of BN, the maximum $R_1$ value for the bond contribution to contrast without regard to noise was only 1.2%, which compares unfavorably to the maximum value of 32.9% for Forsterite. Despite this experimental challenge, local bonding effects have recently been resolved for monolayer BN in HREM images [29] wherein bonding contributions were confirmed both experimentally and computationally to contribute as much as 10% to the single pixel contrast at the nitrogen site. The $R_1$ value computed from the DFT and IAM curves in Figure 4 of [29] is 0.9%, which is consistent with the maximum simulated value of 1.2% presented herein, and represents the spatially averaged contrast contribution over the entire image rather than a localized maximum. Moderately thick samples should offer an even more robust route to real space observation of bonding effects. The full series of $R_{1,bond}$ vs defocus for monolayer BN and 20 nm Forsterite are shown in Figure 13.

## 4. Discussion

It has been shown through analysis of simulated high resolution electron microscopy images that bonding perturbations to the electrostatic potential of a crystal give rise to measurable effects in real space. The magnitude of this effect is highly dependent upon the sample thickness and the microscope conditions under which the images are recorded, but may contribute as much as 30% to the total image contrast. The primary contribution to image perturbations due to bonding in Forsterite comes from the more covalent Si-O bonds rather than the more ionic Mg-O contribution. Moreover, the relatively low spatial frequency of the observed bonding effects (0.2-0.4 $Å^{-1}$), and simulated data for a conventional electron microscope both imply that aberration-corrected microscopy may not in principle be necessary to observe these effects experimentally. For some materials, aberration correction may reduce the observability of bonding effects due to greater damping of the low frequency portion of the phase contrast transfer function. The primary barrier to imaging bonding electrons is not the absolute resolution of the image, but rather the maximum signal to noise ratio achievable.

Analysis of the simulated images and restored exit waves suggests that relatively thick samples (~20-25 nm) would be better for the measurement of bonding effects than the thin (<10nm) samples traditionally used for HREM imaging due to dynamical amplification of bonding contrast. However, it is important to note that the exit wave restorations reported here were carried out using a linear restoring filter. At a thickness of 25nm, approximately 33% of the intensity has been scattered away from the (0,0) reflection and hence the use of a non-linear code may be more appropriate. At very large sample thicknesses, the image contrast may also be substantially affected by anomalous absorption effects not included in the simulations presented herein. However the 25nm optimal thickness for bonding observation is only 60% of the 41nm extinction length for Forsterite in the [010] orientation, which will partially mitigate this effect.

We generally expect bonding to be most observable for lighter element structures, such as Forsterite, due to the large fraction of electrons that participate in bonding. However, because electrons scatter from the electrostatic potential (rather than the square of the local charge density as in x-rays), screening of the positive core will patially mitigate the Z-dependence of bond sensitivity. Therefore, even for heavier elements, small perturbations to the bonding charge density can substantially affect the electrostatic potential due to modifications to the screening of the positive atomic cores. Electron diffraction measurements of bond density have been accomplished for $YBa_2Cu_3O_7$ and $Bi_2Sr_2CaCu_2O_{8+[delta]}$ [30] and other high-Z materials, which implies that other electron scattering techniques, such as HREM imaging, may be sensitive to bonding in these high-Z structures as well. Due to the dynamical amplification of bonding contrast, it is not only the atomic number which is important, but also the repeat distance in the projected column which will determine the energy of the channeling eigenstates.

Work is currently underway to recover the exit waves from simulated focal series using a non-linear imaging code. The relatively strong thickness dependence of dynamical amplification on the bonding contrast suggests that proper determination of sample thickness will be crucial in deconvoluting differences in a figure of merit due to subtle bonding effects from deviations of the true sample thickness in an experimental dataset. Furthermore, the finding that the modulus of the exit wave is more sensitive to bonding than the phase suggests that the experimental result may be quite sensitive to small misorientations of the crystal from the zone axis, which may produce asymmetric contrast that could be incorrectly interpreted as bonding perturbations. Future experimental measurements of bonding in real space may also supplement band gap and lattice parameter measurements used to benchmark the increasing number of DFT functionals aimed at solving the computational problems of strongly correlated materials.

**Aknowledgements**

This work was supported in part by the NSF Materials World Network program through DMR-0710643 (JC and LDM), and EPSRC grant EP/F028784/1 (AIK, JSK).

**Figures**

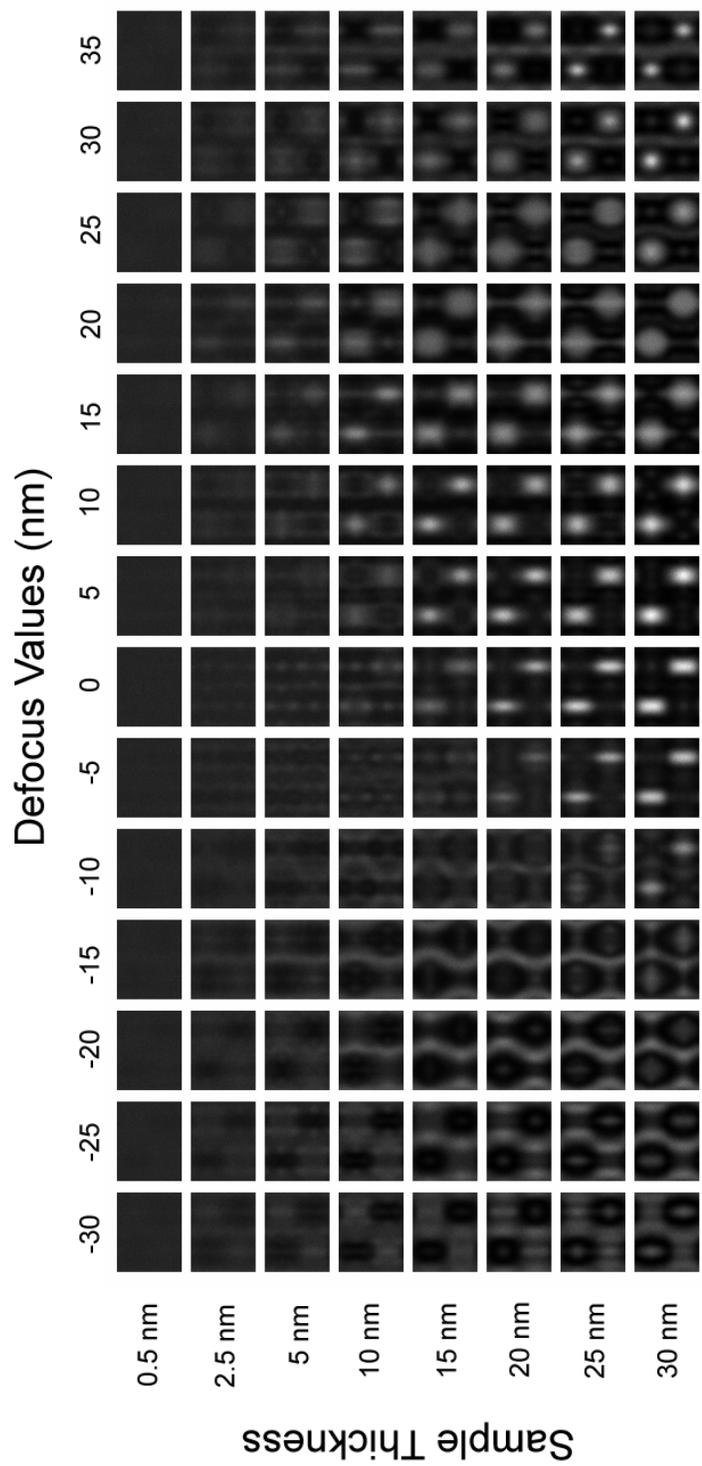

Figure 1: Collage of simulated images of Forsterite [010] using the CD potential for the JEOL-2200MCO microscope with $C_3 = -0.005$ mm. 10% Poisson noise was added to each image. Thickness and defocus values were as indicated. The size of each panel of the collage is 4.75 Å x 5.98 Å.

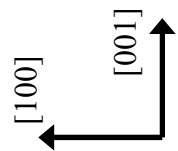

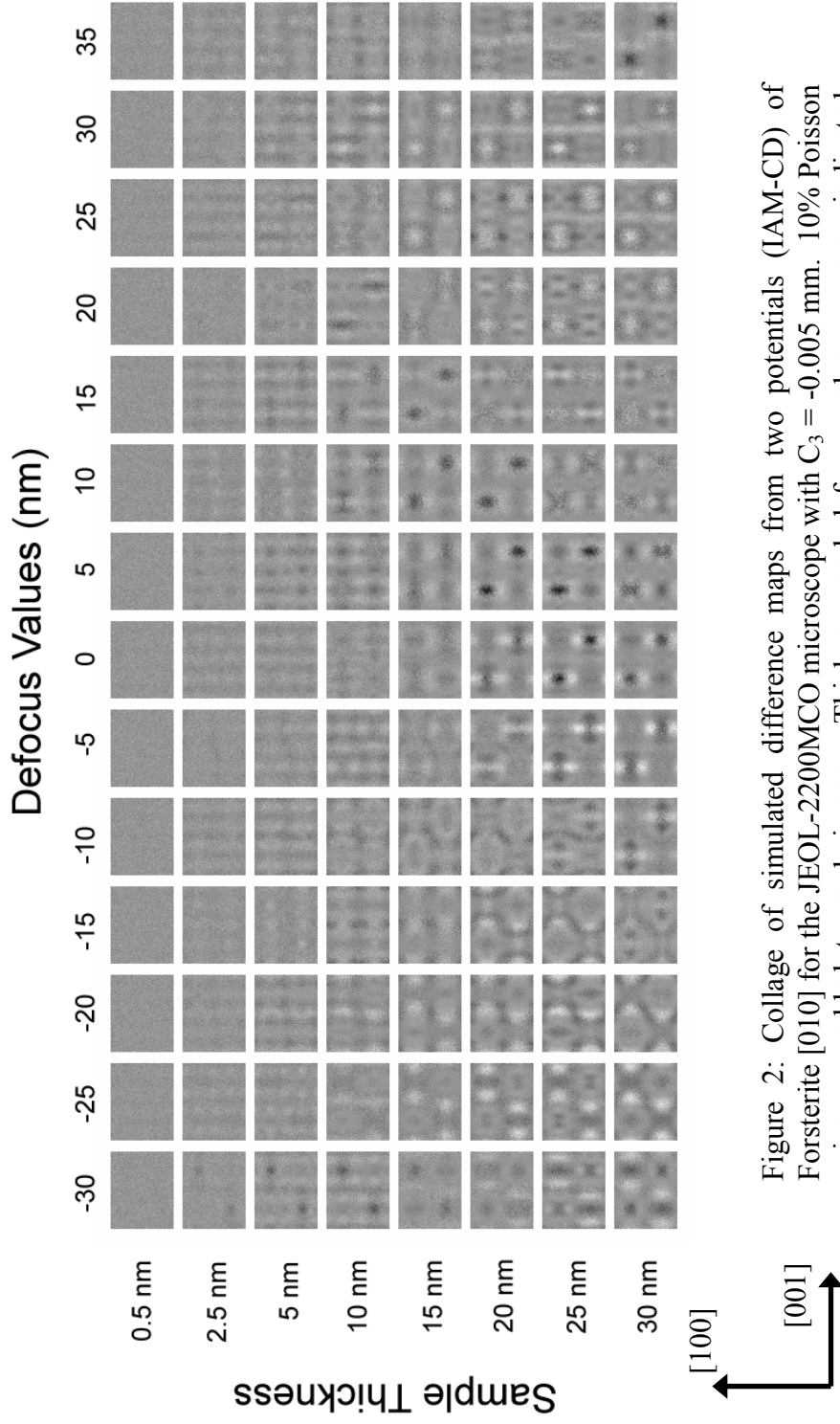

Figure 2: Collage of simulated difference maps from two potentials (IAM-CD) of Forsterite [010] for the JEOL-2200MCO microscope with $C_3 = -0.005$ mm. 10% Poisson noise was added to each image. Thickness and defocus values were as indicated. Contrast scaled to be 5x greater than Figure 1. The size of each panel of the collage is 4.75 Å x 5.98 Å.

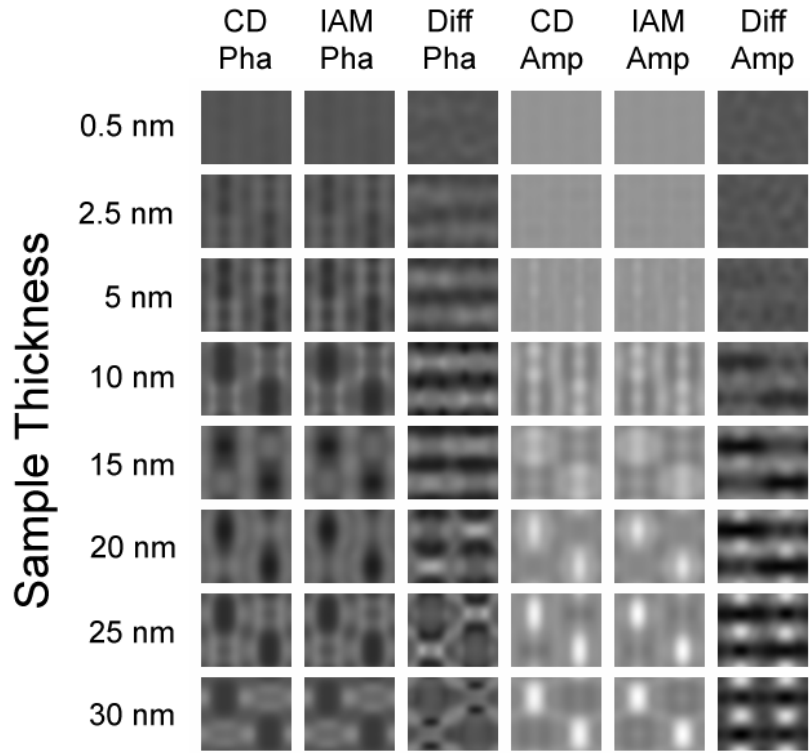

Figure 3: Modulus and phase of exit waves restored from focal series of noisy images with $C_3$ = -0.005 mm. Contrast in difference maps is increased 10x.

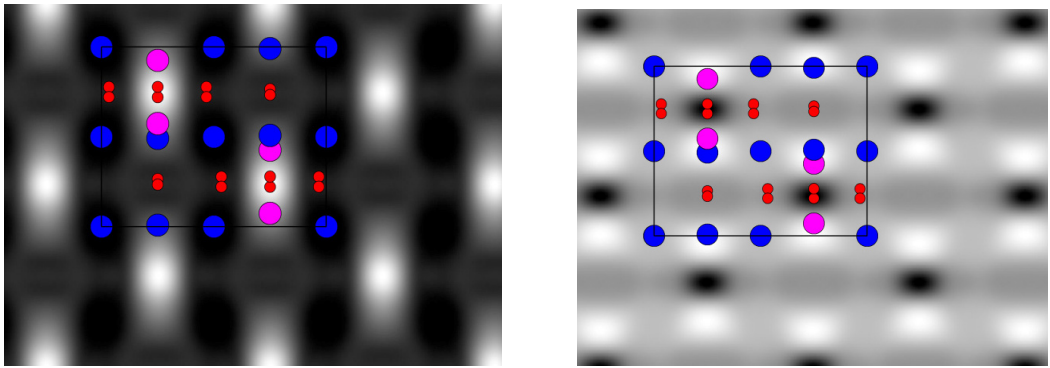

Figure 4: Simulated HREM images of Forsterite [010] projection, 15nm thick, +10nm defocus, -0.005mm $C_3$  a) individual CD image  b) CD-IAM difference map.  Silicon atoms pink, magnesium atoms blue, oxygen atoms red.

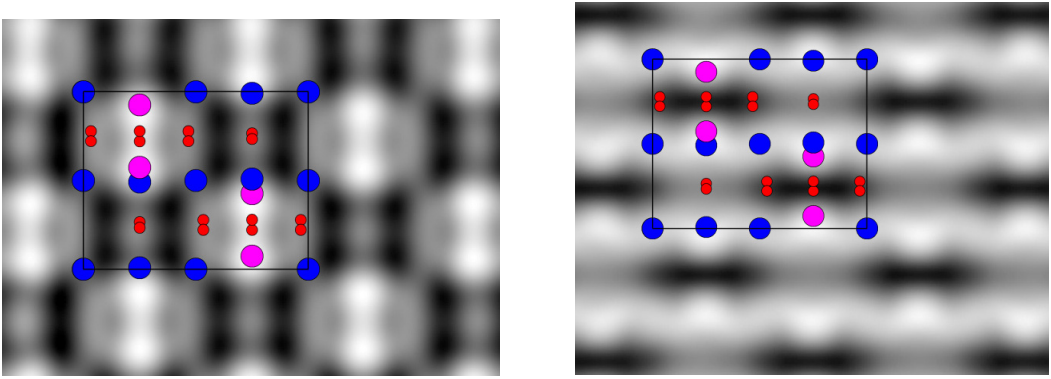

Figure 5: Modului of exit waves restored from simulated HREM images of Forsterite [010] projection, 15nm thick  a) restored CD modulus   b) CD-IAM difference map of modulus. Silicon atoms pink, magnesium atoms blue, oxygen atoms red.

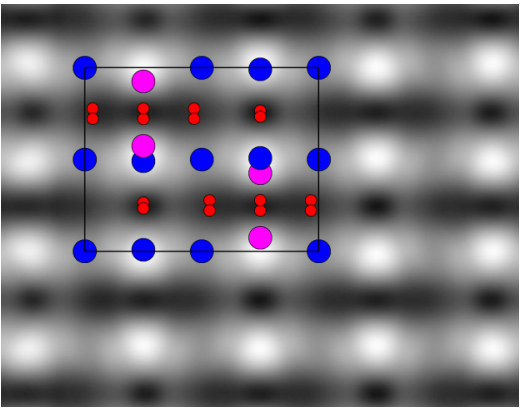

Figure 6: CD-IAM difference map of exit wave modulus restored from simulated HREM images of Forsterite [010] projection, 25nm thick. Silicon atoms pink, magnesium atoms blue, oxygen atoms red.

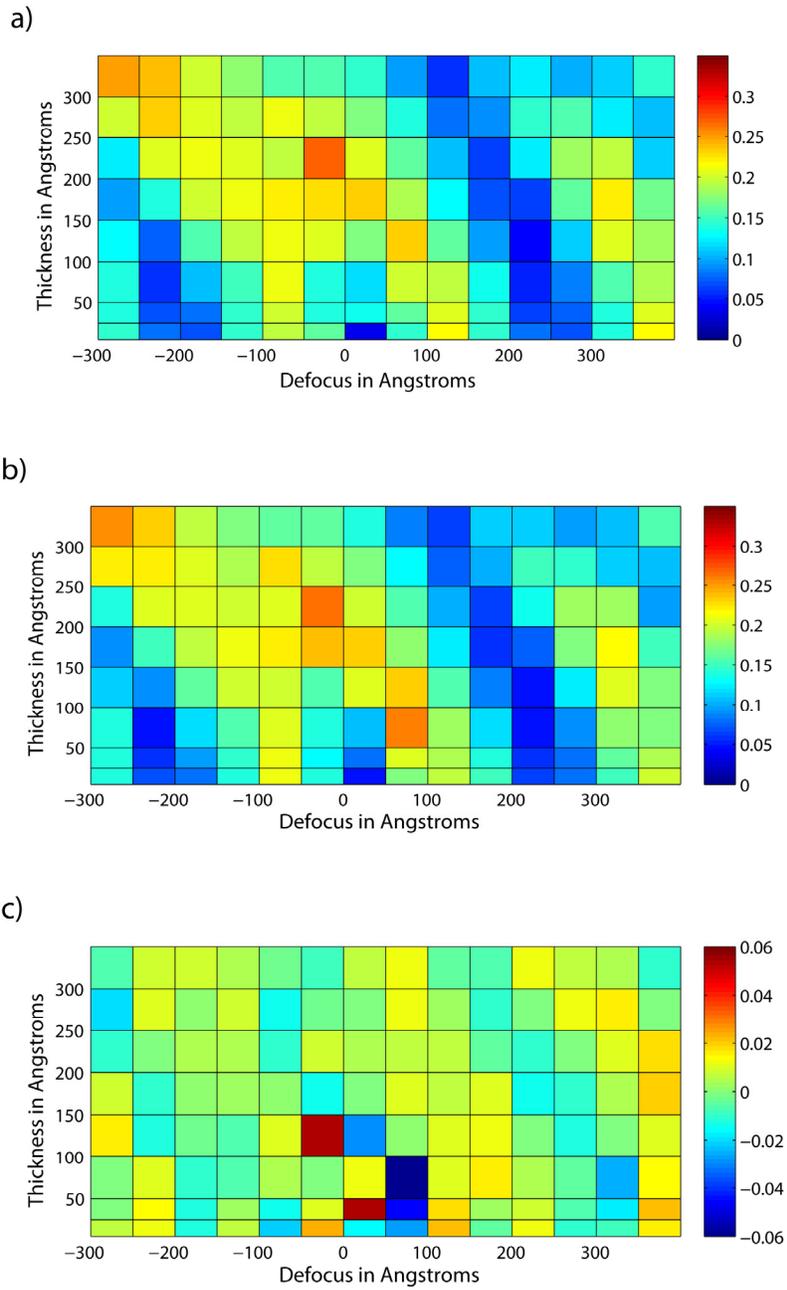

Figure 7: $R_\sigma$ metric for bond observability for a range of thickness and defocus values with a) $C_3$ = -0.005 mm, b) $C_3$ = +0.005mm, and c) the difference between $+C_3$ and $-C_3$ conditions

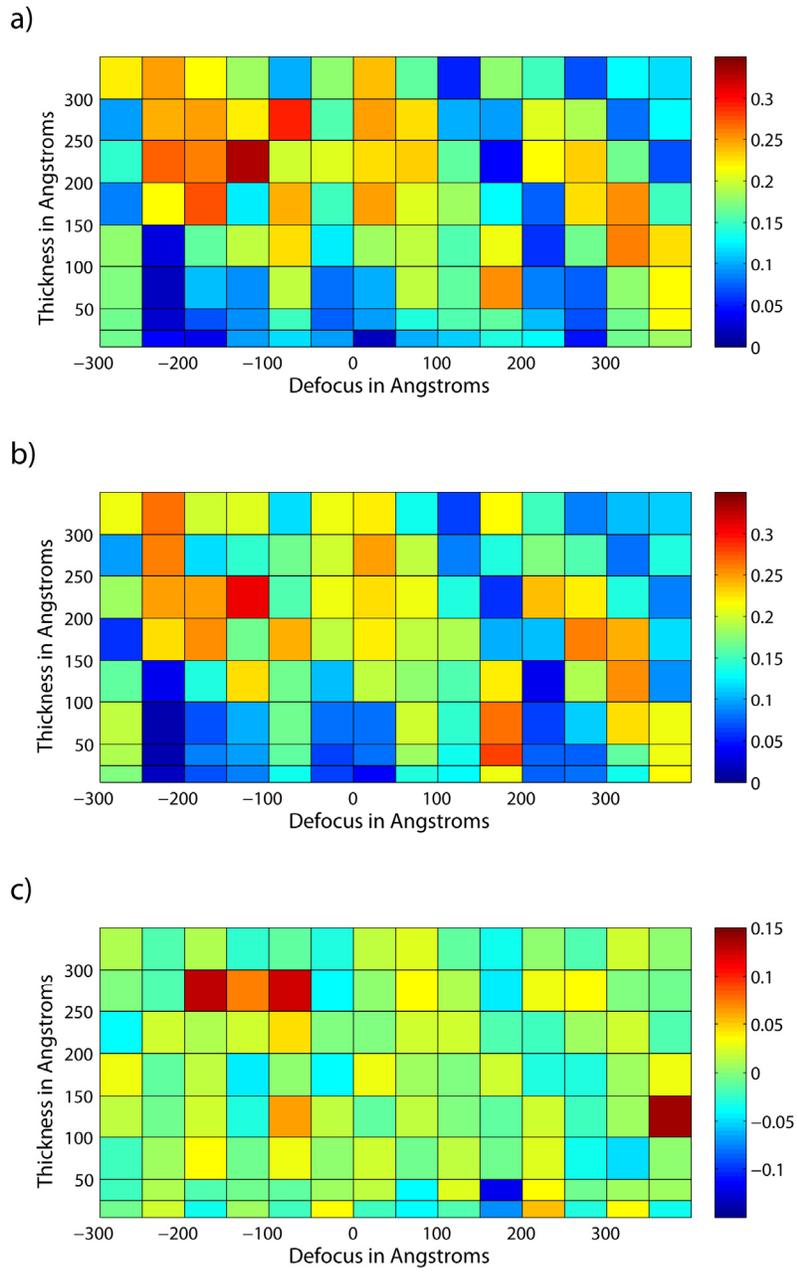

Figure 8: $R_1$ metric for bond contribution for a range of thickness and defocus values with a) $C_3$ = -0.005 mm, b) $C_3$ = +0.005mm, and c) the difference between $+C_3$ and $-C_3$ conditions. The maximum $R_1$ value of 32.9%, occurs for $C_3$ = -0.005 mm at 20 nm thickness and -15 nm defocus.

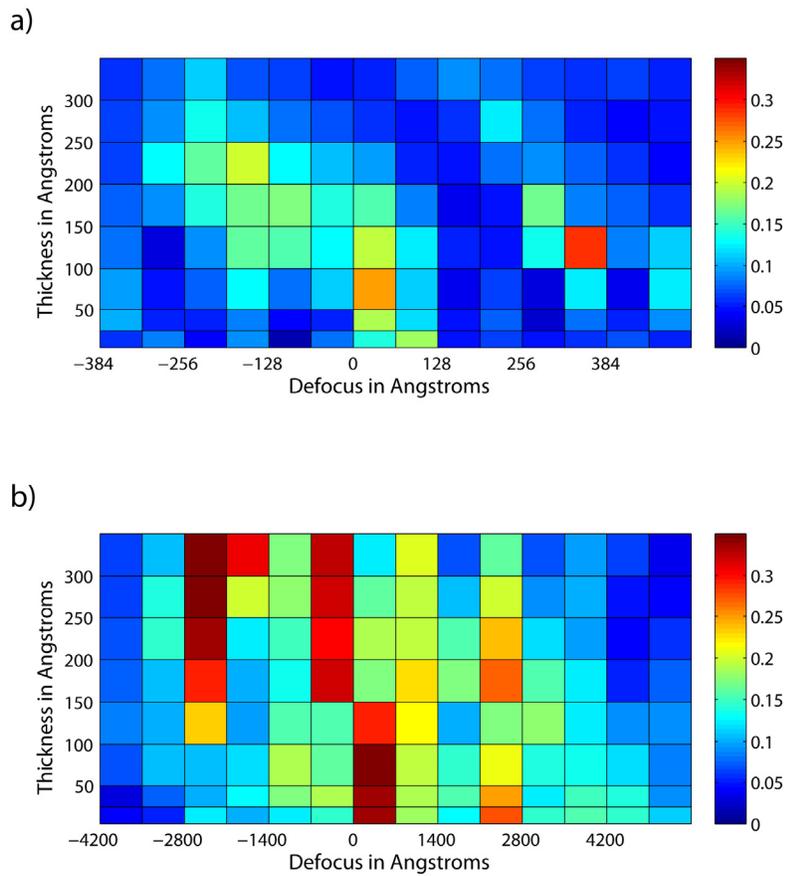

Figure 9: $R_1$ metric for bond contribution at a) 80 kV for a range of thickness and defocus values with C3 = -0.005 mm b) 200 kV for a range of thickness and defocus values with C3 = 1.00 mm. Defocus values correspond to the same fractions of Scherzer defocus as seen in Figure 8.

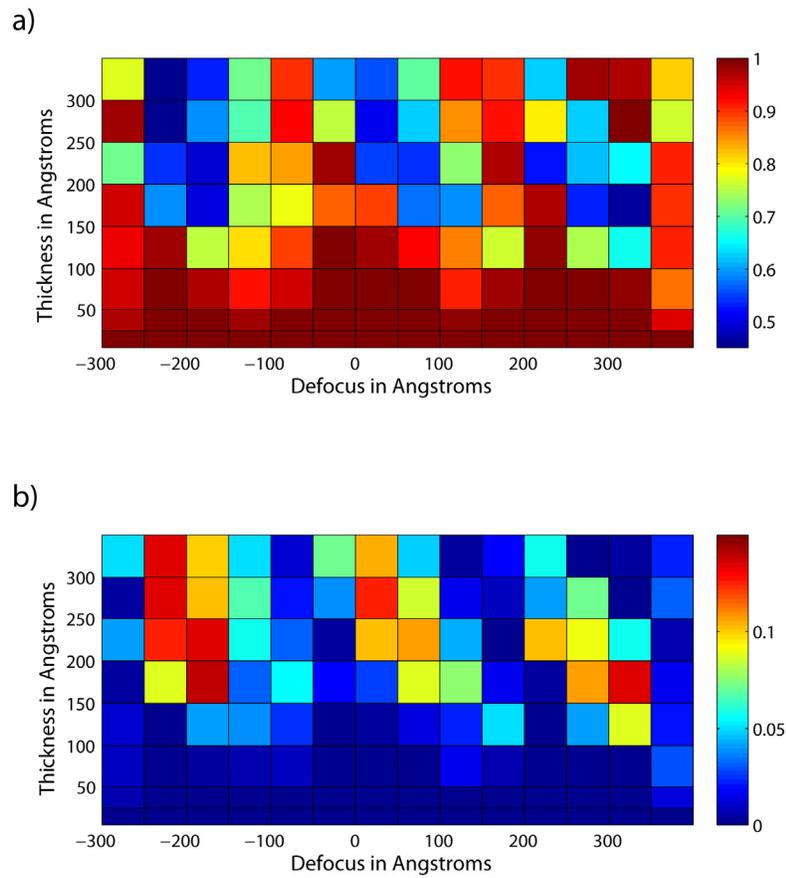

Figure 10: Thickness and defocus map of a) $R_1(\textit{clean,noisy})$, smaller is better and b) $R_{1,obs}$, larger is better. The maximum $R_{1,obs}$ value of 13.8%, occurs at 15 nm thickness and -15 nm defocus.

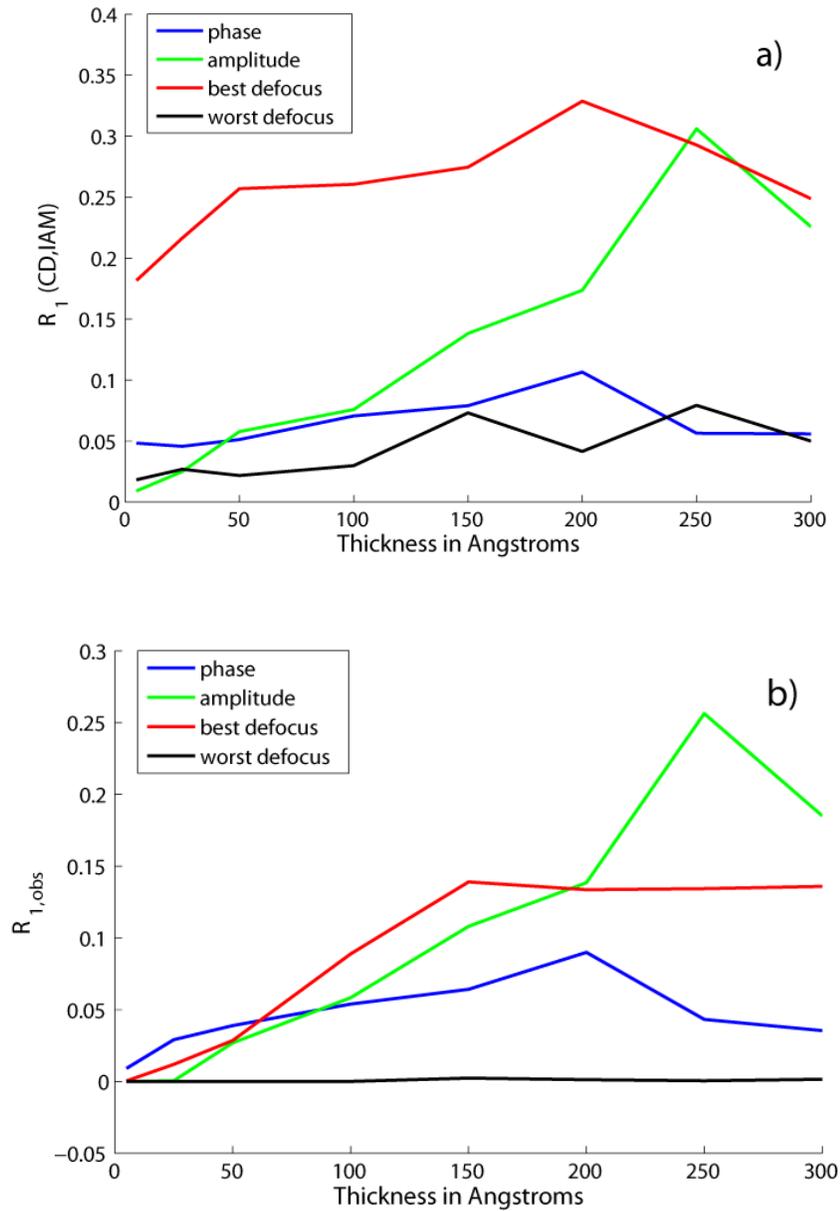

Figure 11: Figures of merit for the modulus and phase of restored exit waves from simulated focal series. a) $R_1(CD, IAM)$ metric without noise b) $R_{1,obs}$ metric including noise. For purposes of comparison, the best and worst achievable bonding sensitivity for single images at each thickness (from Figure 8b) are also shown.

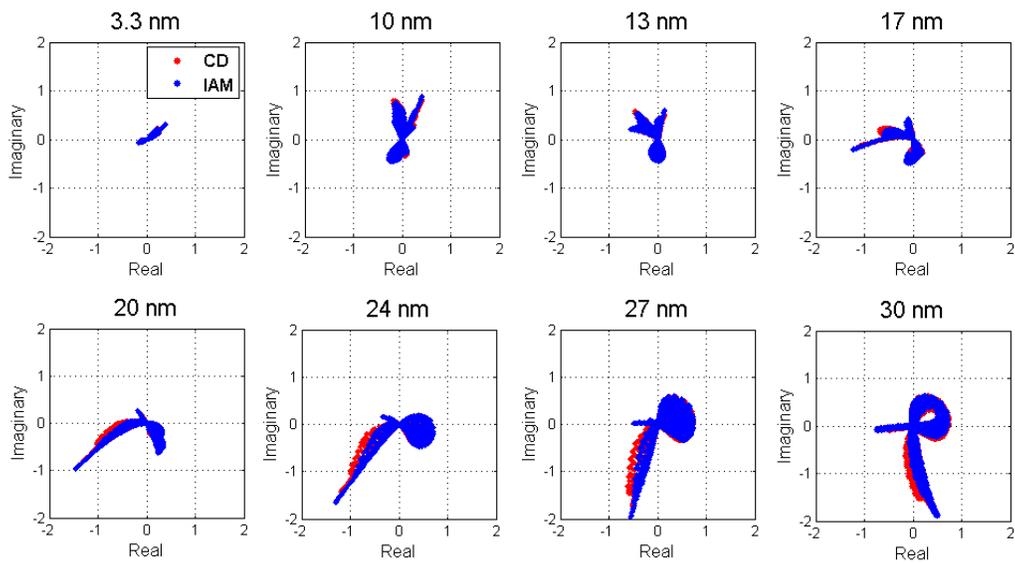

Figure 12: Argand plots of Ψ(r)-1 for Forsterite projected along the [010] direction at thickneses from 33 to 300 Å. Ψ(r) calculated from both isolated atom (IAM) and full charge density potentials (CD) using multislice. All figures are shown to scale.

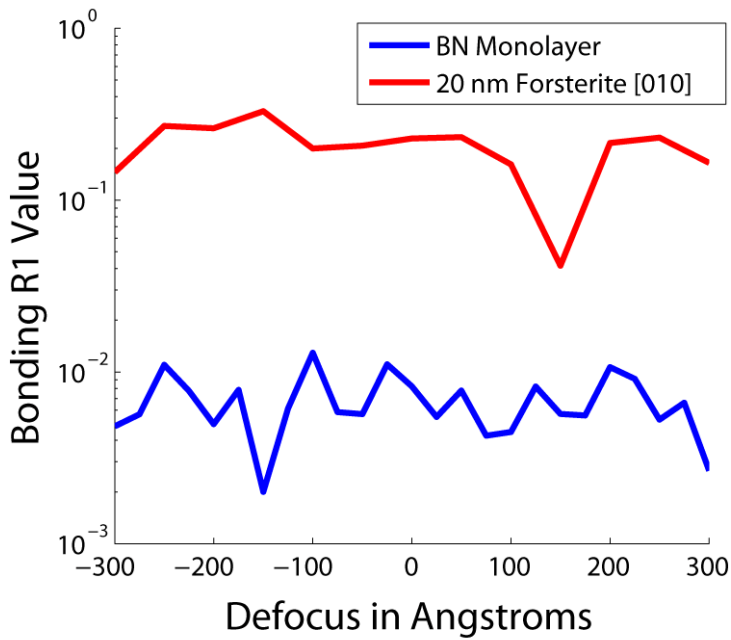

Figure 13: Comparison of the $R_1$ metric for bond contribution in monolayer BN and 20 nm Forsterite [010] at a range of defocus values. Note logarithmic scaling of the ordinate.